\documentclass[sigplan,screen,sigconf]{acmart}

\usepackage{soul} 
\usepackage{cleveref}
\usepackage{xspace}

\AtBeginDocument{%
  \providecommand\BibTeX{{%
    \normalfont B\kern-0.5em{\scshape i\kern-0.25em b}\kern-0.8em\TeX}}}

\setcopyright{rightsretained}
\acmDOI{}
\acmISBN{}
\usepackage{circuitikz}

\acmConference[LATTE '24]
{4th Workshop on Languages, Tools, and Techniques for Accelerator Design}
{April 28, 2024}
{San Diego, CA, USA}

\newcommand{\egraph}{\mbox{e-graph}\xspace}

\newcommand{\Eqsat}{Equality saturation\xspace}
\newcommand{\eqsat}{equality saturation\xspace}

\begin{document}

\title{There and Back Again: \\ A Netlist's Tale with Much Egraphin'}

\author{
Gus Henry Smith\textsuperscript{\textdagger}, 
Zachary D. Sisco\textsuperscript{\textdaggerdbl},
Thanawat Techaumnuaiwit\textsuperscript{\textdaggerdbl}, 
Jingtao Xia\textsuperscript{\textdaggerdbl}, 
Vishal Canumalla\textsuperscript{\textdagger}, 
Andrew Cheung\textsuperscript{\textdagger}, 
Zachary Tatlock\textsuperscript{\textdagger},
Chandrakana Nandi\textsuperscript{\S}, 
Jonathan Balkind\textsuperscript{\textdaggerdbl} \\
\small \textsuperscript{\textdagger} University of Washington \
\small \textsuperscript{\textdaggerdbl} University of California, Santa Barbara \
\small \textsuperscript{\S} Certora, Inc. \\
\small \texttt{ \{gussmith, vishalc, acheung8, ztatlock\}@cs.washington.edu} \ 
\small \texttt{\{zsisco, thanawat, jingtaoxia, jbalkind\}@ucsb.edu} \ 
\small \texttt{chandra@certora.com}
}

\renewcommand{\shortauthors}{Smith, et al.}

\begin{abstract}
EDA toolchains are notoriously 
  unpredictable, incomplete, and error-prone;
  the generally-accepted remedy
  has been to re-imagine EDA tasks
  as compilation problems.
However, any compiler framework
  we apply
  must be prepared to handle
  the wide range of EDA tasks,
  including not only compilation tasks
  like technology mapping and optimization
  (the \textit{``there''} in our title),
  but also decompilation tasks like loop rerolling
  (the \textit{``back again''}).
In this paper,
  we advocate for
  \textit{\eqsat}---%
  a term rewriting framework---%
  as the framework of choice
  when building hardware toolchains.
Through a series of case studies, 
  we show how
  the needs of EDA tasks
  line up conspicuously well
  with the features
  \eqsat provides.
\end{abstract}

\maketitle

\section{Introduction}

Hardware development toolchains
  are notorious
  for their
  unpredictability~\cite{nigam2020predictable},
  incompleteness~\cite{smith2024fpga},
  and incorrectness~\cite{herklotz2020finding}.
These issues
  stem from the fact
  that most common toolchains
  do not treat EDA tasks
  as compilation problems,
  and instead often
  use ad hoc,
  unprincipled approaches
  to solving each problem.
Existing projects
  such as MLIR CIRCT~\cite{circt},
  LLHD~\cite{schuiki2020llhd},
  and Calyx~\cite{nigam2021compiler}
  have made great strides
  towards reframing
  and restructuring
  hardware design tools
  using consistent 
  compiler frameworks.

Finding an appropriate
  compiler framework
  is difficult, as
  the EDA tasks that
  must be supported
  are diverse.
For example, any
  framework should certainly
  be able to capture
  all standard optimization tasks,
  such as register retiming,
  pipelining, and common
  subexpression elimination.
However, another essential
  task beyond 
  standard optimization is 
  \textit{technology mapping}---the
  process of implementing
  a high-level design specification
  using the actual hardware primitives
  available on the target FPGA or ASIC
  process. 
To make things even more
  complicated,
  EDA tasks are not
  always ``moving forward'':
  recent work has 
  established \textit{hardware decompilation}
  as a valuable tool for design
  tasks such as speeding 
  up netlist simulations~\cite{Sisco2023}.
Thus, an ideal compiler framework
  must also be able to easily
  break and lower between levels
  of abstraction.


\textbf{\Eqsat}~\cite{peggy}
  is a compiler framework
  which has already proven its prowess
  in all of these tasks.
\Eqsat is a non-destructive
  term rewriting technique
  that uses the \textbf{\egraph data structure}~\cite{nelson, nieuwenhuis05} 
  to compactly store potentially infinitely many
  equivalent terms.
Recent work~\cite{egg, egglog}
  has developed fast and extensible libraries for efficient
  \eqsat.
Previous work has shown \eqsat's ability to implement
  decompilation~\cite{szalinski},
  procedural abstraction~\cite{babble},
  optimization~\cite{spores,
    laddad2023optimizing, 
    zhao2023automatic,
    thomas2024automatic, 
    wang2023infinity,
    matsumura2023acc},
  and 
  mapping~\cite{Glenside,3LA, Diospyros}.

 
  

\textbf{In this position paper,
  we advocate for the extensive application of
  \eqsat
  to EDA tasks.}
In fact, \eqsat
  has already shown early promise
  in being applied to various EDA tasks,
  including
  RTL optimization~\cite{pi2023esfo,ho2023wolfex},
  HLS optimization~\cite{Seer},
  multiplier optimization~\cite{wanna2023multiplier,ustun2022impress},
  and repurposing CGRAs~\cite{woodruff2023rewriting}.
Ustun et al.\ also
  argue for \eqsat
  in datapath synthesis and optimization~\cite{Ustun}.
We make a larger claim in this paper
  that \eqsat has value
  beyond optimization
  in EDA tasks up and down the stack.

We now present four case studies
  highlighting different properties
  of \eqsat that makes it attractive
  at different stages
  in the hardware design workflow,
  from technology mapping
  to circuit-level analyses
  such as retiming and decompilation.
These case studies 
    demonstrate that the operational
    semantics of the various
    stages of the hardware design 
    workflow are intuitively represented as
    rewrite rules.
Each case study will explore
    the different properties
    of \eqsat that makes it
    attractive for implementing
    the different hardware passes.

\section{Case Study: Hardware Loop Rerolling}
\label{sec:loop-rerolling}
Recent work considers the problem of hardware loop rerolling,
    that is, identifying repeated sequences 
    of logic in a netlist and rerolling 
    them into loops in higher-level HDL code~\cite{Sisco2023}.
This research fits into the larger problem 
    of \textit{hardware decompilation}, 
    which lifts netlists to HDL code to help 
    with design and analysis tasks. 
Loop rerolling for hardware decompilation 
    uses a sketch-guided program synthesis 
    technique to synthesize rerolled loops~\cite{Sisco2023}.
However, this technique scales poorly due to its 
  reliance on SMT solvers to fill in the loop sketches.

Our in-progress work considers hardware loop rerolling 
  through the lens of rewriting.
Consider the following illustration
  of a rewrite rule which 
  identifies a repeated logic block \texttt{G} 
  and rewrites it into a \texttt{for}-loop with \texttt{G} 
  parameterized over the loop variable \texttt{i}:

\begin{figure}[!ht]
\centering
\resizebox{0.48\textwidth}{!}{%
\begin{circuitikz}
\tikzstyle{every node}=[font=\Huge]
\draw [ fill={rgb,255:red,222; green,222; blue,222} ] (1.5,23.75) rectangle  node {\Huge \texttt{G}} (4,22.25);
\draw [ fill={rgb,255:red,222; green,222; blue,222} ] (4.25,23.75) rectangle  node {\Huge \texttt{G}} (6.75,22.25);
\draw [ fill={rgb,255:red,222; green,222; blue,222} ] (8.25,23.75) rectangle  node {\Huge \texttt{G}} (10.75,22.25);
\draw [->, >=Stealth] (2.25,24.25) -- (2.25,23.75);
\draw [->, >=Stealth] (3.25,24.25) -- (3.25,23.75);
\draw [->, >=Stealth] (5,24.25) -- (5,23.75);
\draw [->, >=Stealth] (6,24.25) -- (6,23.75);
\draw [->, >=Stealth] (9,24.25) -- (9,23.75);
\draw [->, >=Stealth] (10,24.25) -- (10,23.75);
\draw [->, >=Stealth] (2.75,22.25) -- (2.75,21.75);
\draw [->, >=Stealth] (5.5,22.25) -- (5.5,21.75);
\draw [->, >=Stealth] (9.5,22.25) -- (9.5,21.75);
\draw  (14,24.5) rectangle (18.75,21.75);
\node [font=\LARGE] at (2.25,24.5) {$a_0$};
\node [font=\LARGE] at (3.25,24.5) {$b_0$};
\node [font=\LARGE] at (5,24.5) {$a_1$};
\node [font=\LARGE] at (6,24.5) {$b_1$};
\node [font=\LARGE] at (9,24.5) {$a_n$};
\node [font=\LARGE] at (10,24.5) {$b_n$};
\node [font=\LARGE] at (2.75,21.5) {$c_0$};
\node [font=\LARGE] at (5.5,21.5) {$c_1$};
\node [font=\LARGE] at (9.5,21.5) {$c_n$};
\node [font=\Huge] at (11.75,23.25) {};
\node [font=\huge] at (16.35,23.9) {\texttt{for i = 0 .. n}};
\draw [ fill={rgb,255:red,222; green,222; blue,222} ] (15.5,23.25) rectangle  node {\Huge \texttt{G}$_i$} (17.25,22);
\node [font=\Huge] at (12.75,23.75) {$a_{[0..n]}$};
\node [font=\Huge] at (12.75,22.25) {$b_{[0..n]}$};
\node [font=\Huge] at (20.25,23) {$c_{[0..n]}$};
\draw [->, >=Stealth] (15,23) -- (15.5,23);
\draw [->, >=Stealth] (15,22.25) -- (15.5,22.25);
\draw [->, >=Stealth] (17.25,22.5) -- (18,22.5);
\node [font=\LARGE] at (14.75,23) {$a_i$};
\node [font=\LARGE] at (14.75,22.25) {$b_i$};
\node [font=\LARGE] at (18.25,22.5) {$c_i$};
\draw [->, >=Stealth] (13.5,23.75) -- (14,23.75);
\draw [->, >=Stealth] (13.5,22.25) -- (14,22.25);
\draw [->, >=Stealth] (18.75,23) -- (19.5,23);
\draw [, dashed] (1.25,25.25) rectangle  (11,21);
\draw [, dashed] (12,24.75) rectangle  (21,21.5);
\node [font=\Huge] at (11.5,23) {$\rightsquigarrow$};
\node [font=\Huge] at (7.5,23) {...};
\end{circuitikz}
}%
\end{figure}


While in this example, the indices $a$ and $b$ appear in
  monotonically increasing order on the logic blocks,
  this may not always be the case, which would make it
  harder to infer the closed form for the \texttt{for}-loop.
More generally, loop rerolling is particularly
    challenging when the initial, 
    unrolled program does not expose any high-level structure, 
    i.e., the repetitive patterns of the program are obfuscated.
Prior work shows that \eqsat can be used to discover this latent structure 
    by applying carefully designed rewrite rules~\cite{szalinski}.
We envision scaling hardware loop rerolling 
    by leveraging similar techniques.


%
%
%
%

\section{Case Study: Standard Library Component Identification}
\label{sec:stl}
This case study is about
   finding components from a hardware standard library 
   within a compiled artifact such as a netlist.
The compiler optimizes the component in ways that 
  using a sub-graph-isomorphism algorithm for identification will fail, 
  and, for large enough designs, will not scale.
An \egraph solves these problems in two ways:
  (1) it allows us to explore semantically equivalent 
    versions of the same design to 
    find the one where we can extract the standard 
    library component and
  (2) it allows sub-graphs to be extracted out more 
    efficiently due to the internal union-find structure.
For standard library component identification, 
    we can directly take the standard library component we are looking
    for and turn it into a rewrite rule within 
    the \verb|egglog| \eqsat engine~\cite{egglog}.
For example, here is an illustration of the rewrite rule for a half adder:

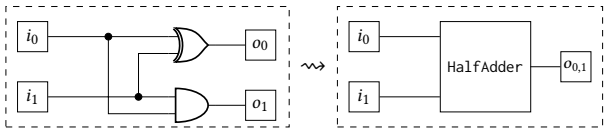
\begin{figure}[!ht]
\centering
\resizebox{0.45\textwidth}{!}{%
\begin{circuitikz}
\tikzset{input/.style={draw , minimum height=0.5cm,  minimum width=.5cm,  anchor=east}}

\tikzset{output/.style={draw , minimum height=0.5cm,  minimum width=.5cm,  anchor=west}}

\draw (0,1) node[input, label=center:$i_0$](ia) {} to 
(2,1) node[xor port, anchor=in 1, scale=0.5] (x) {}
(x.in 2) -- ++(-0.5,0) to[short, -*] (1.5,0)
(x.out) to ++(0.5,0) node[output,label=center:$o_0$]{};
\draw (0, 0) node[input, label=center:$i_1$](ib) {} to
(2,0) node[and port, anchor=in 1, scale=0.5] (a) {}
(a.in 2) -- ++(-1,0) to[short, -*] (1,1)
(a.out) to ++(0.5,0) node[output, label=center:$o_1$]{};

\node [font=\Large] at (4.4,0.5) {$\rightsquigarrow$};

\draw [, dashed] (-0.7,-0.5) rectangle  (4.0,1.5);
\draw [, dashed] (4.8,-0.5) rectangle  (9.2,1.5);

\draw (5.5,1) node[input, label=center:$i_0$] {} -- ++(1,0);
\tikzset{halfadd/.style={draw , minimum height=1.5cm,  minimum width=1.5cm,  anchor=west}}
\draw (6.5,0.5) node[halfadd, label=center:{\small \texttt{HalfAdder}}] {};
\draw (8.5, 0.5) node[ output, label=center:{\small $o_{0,1}$}]{} -- ++(-0.5,0); 
\draw (5.5,0) node[input, label=center:$i_1$] {} -- ++(1,0);
\end{circuitikz}
}%
\end{figure}


\noindent Within this rewrite rule, $i_0$ and $i_1$ are \textit{any arbitrary circuits}.
\Eqsat runs this rewrite rule (along with standard rules for Boolean algebra) on a 
    larger design which pattern-matches parts of the design with
    the half-adder definition---rewriting that definition into 
    an abstract half-adder component.

Challenges with this approach include
  matching on components
  where the compiler optimized away parts of the module
  or fused two modules together which share resources.
Anti-unification techniques,
  as presented in babble~\cite{babble},
  can help with the problem of partial matching.
Further,
  a generalized problem of standard library identification
  is \textit{procedural abstraction},
  finding repeated instances of a procedure
  where there is no standard library as reference to match against.

\section{Case Study: Scaling Technology Mapping via Library Learning}

Our previous work Lakeroad~\cite{smith2024fpga}
  demonstrates how
  the process of FPGA \textit{technology mapping}---%
  converting a high-level hardware design description
  into an implementation using FPGA-specific primitives---%
  can be vastly improved
  via 
  program synthesis.
However,
  program synthesis is known
  to face scaling issues.
Meanwhile, the process of technology mapping
  must scale to potentially
  massive hardware designs.

With \eqsat,
  we can scale these state-of-the-art
  technology mapping techniques
  via the application of
  \textit{library learning}~\cite{babble}.
Library learning
  is the process of finding
  abstractions
  commonly used
  throughout a corpus of code---%
  in our setting,
  finding hardware modules
  used repeatedly
  within a larger design.
Within the \eqsat framework,
  library learning
  can be expressed simply as a rewrite
  which converts an expression
  into an abstracted module applied
  to a list of concrete inputs:

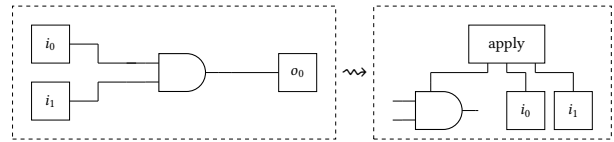
\begin{figure}[!ht]
\centering
\resizebox{.45\textwidth}{!}{%
\begin{circuitikz}
\tikzstyle{every node}=[font=\Huge]
\draw (3.75,9.75) to[short] (4.25,9.75);
\draw (3.75,9.25) to[short] (4.25,9.25);
\draw (4.25,9.75) node[ieeestd and port, anchor=in 1, scale=0.89](port){} (port.out) to[short] (6.5,9.5);
\draw  (12.75,10.75) rectangle  node {\LARGE apply} (14.75,9.75);
\draw  (1.25,10.75) rectangle  node {\LARGE $i_0$} (2.25,9.75);
\draw  (1.25,9.25) rectangle  node {\LARGE $i_1$} (2.25,8.25);
\draw [short] (4,9.75) -- (3,9.75);
\draw [short] (3,9.75) -- (3,10.25);
\draw [short] (3,10.25) -- (2.25,10.25);
\draw [short] (3.75,9.25) -- (3,9.25);
\draw [short] (3,9.25) -- (3,8.75);
\draw [short] (3,8.75) -- (2.25,8.75);
\draw  (7.75,10) rectangle  node {\LARGE $o_0$} (8.75,9);
\draw [short] (6.5,9.5) -- (7.75,9.5);
\draw [, dashed] (0.75,11.25) rectangle  (9.25,7.75);
\draw (10.75,8.75) to[short] (11,8.75);
\draw (10.75,8.25) to[short] (11,8.25);
\draw (11,8.75) node[ieeestd and port, anchor=in 1, scale=0.89](port){} (port.out) to[short] (13,8.5);
\draw  (13.75,9) rectangle  node {\LARGE $i_0$} (14.75,8);
\draw  (15,9) rectangle  node {\LARGE $i_1$} (16,8);
\draw [short] (13.25,9.75) -- (13.25,9.5);
\draw [short] (13.25,9.5) -- (11.75,9.5);
\draw [short] (11.75,9.5) -- (11.75,9.25);
\draw [short] (13.75,9.75) -- (13.75,9.5);
\draw [short] (13.75,9.5) -- (14.25,9.5);
\draw [short] (14.25,9.5) -- (14.25,9);
\draw [short] (14.5,9.75) -- (14.5,9.5);
\draw [short] (14.5,9.5) -- (15.5,9.5);
\draw [short] (15.5,9.5) -- (15.5,9);
\draw [short] (11.75,9.25) -- (11.75,9);
\draw [, dashed] (10.25,11.25) rectangle  (16.5,7.75);
\node [font=\Huge] at (9.75,9.5) {$\rightsquigarrow$};
\end{circuitikz}
}%
\end{figure}

\noindent When applied repeatedly across a large design,
  this rewrite will find larger and larger
  abstracted submodules.
By default,
  \eqsat deduplicates identical expressions,
  allowing us to discover submodules which are
  frequently reused across the design.
These abstracted submodules 
  are then perfect candidates for
  program synthesis.
Furthermore, we can use information
  about frequency of appearance
  and other contextual information
  in the \egraph
  to filter and rank candidates.
Thus, \eqsat gives us a path
  towards scaling currently limited
  state-of-the-art techniques
  using simple algebraic rewrites
  and its native deduplication ability.
 
\section{Case Study: Circuit Retiming}

With an algebraic representation of the netlist we form a bidirectional rewrite rule that captures forward and backward retiming:

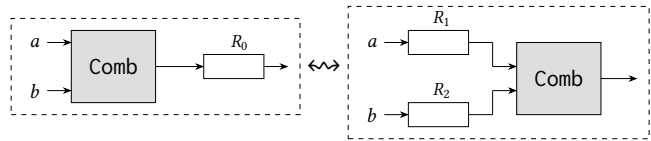
\begin{figure}[!ht]
\centering
\resizebox{0.48\textwidth}{!}{%
\begin{circuitikz}
\tikzstyle{every node}=[font=\Huge]
\draw [ fill={rgb,255:red,222; green,222; blue,222} ] (3,14) rectangle  node {\huge \texttt{Comb}} (4.75,12.5);
\draw [->, >=Stealth] (2.5,13.75) -- (3,13.75);
\draw [->, >=Stealth] (2.5,12.75) -- (3,12.75);
\node [font=\LARGE] at (2.25,13.75) {$a$};
\node [font=\LARGE] at (2.25,12.75) {$b$};
\draw [->, >=Stealth] (4.75,13.25) -- (5.75,13.25);
\draw  (5.75,13.5) rectangle (7,13);
\draw [->, >=Stealth] (7,13.25) -- (7.5,13.25);
\draw  (10,14) rectangle (11.25,13.5);
\draw  (10,12.5) rectangle (11.25,12);
\draw [->, >=Stealth] (9.5,13.75) -- (10,13.75);
\draw [->, >=Stealth] (9.5,12.25) -- (10,12.25);
\draw [ fill={rgb,255:red,222; green,222; blue,222} ] (12.25,13.75) rectangle  node {\huge \texttt{Comb}} (14,12.25);
\draw [->, >=Stealth] (14,13) -- (14.75,13);
\draw [short] (11.25,13.75) -- (11.75,13.75);
\draw [short] (11.75,13.75) -- (11.75,13.25);
\draw [->, >=Stealth] (11.75,13.25) -- (12.25,13.25);
\draw [short] (11.25,12.25) -- (11.75,12.25);
\draw [short] (11.75,12.25) -- (11.75,12.75);
\draw [->, >=Stealth] (11.75,12.75) -- (12.25,12.75);
\draw [, dashed] (1.75,14.25) rectangle  (7.75,12.25);
\draw [, dashed] (8.75,14.5) rectangle  (15,11.75);
\node [font=\LARGE] at (9.25,13.75) {$a$};
\node [font=\LARGE] at (9.25,12.25) {$b$};
\node [font=\Large] at (6.5,13.75) {$R_0$};
\node [font=\Large] at (10.7,14.25) {$R_1$};
\node [font=\Large] at (10.7,12.75) {$R_2$};
\node [font=\Huge] at (8.25,13.25) {$\leftrightsquigarrow$};
\end{circuitikz}
}%
\vspace{-0.20cm}
\end{figure}

\noindent where \texttt{Comb} is a combinational gate.
With only these two rules,
  \eqsat explores
  all possible ways of arranging registers
  in the design through non-destructive rewrites.
Then, we use ILP (Integer Linear Programming) to
  retime the circuit according to a cost function---%
  following prior work
  that effectively uses ILP extraction
  from an \egraph~\cite{spores,Seer}.

The other side of retiming is \textit{undoing} the effects of a retimed cicuit by, for example, moving all registers as close as possible to their source.
This pass is useful for decompilation by moving registers outside of a section of combinational logic to expose latent structure for other analyses such as standard library component identification and loop rerolling (\Cref{sec:loop-rerolling,sec:stl}).

\section{Conclusion}
We present case studies
  demonstrating how \eqsat can be used to improve
  state of the art techniques for mitigating
  four concrete hardware challenges:
  decompilation through loop rerolling,
  library component identification and
  technology mapping through library learning,
  and optimum circuit retiming through efficient
  state space exploration and ILP extraction.
We are already working on some of these topics and
  hope this paper encourages other researchers to
  consider \eqsat as a technique 
  to mitigate
  EDA challenges in the future.



\bibliography{biblio}
\end{document}